\documentclass[reprint,twocolumn,aps,prB,amsmath,amssymb,floatfix,superscriptaddress,longbibliography]{revtex4-1}
\usepackage{graphicx}
\usepackage{epsfig}
\usepackage{bm}
\usepackage{dcolumn}
\usepackage{color}
\usepackage{physics}
\usepackage{float}
\usepackage[colorlinks,urlcolor=blue,citecolor=blue,linkcolor=magenta]{hyperref}
\makeatletter
\newcommand*{\rom}[1]{\expandafter\@slowromancap\romannumeral #1@}
\makeatother
\usepackage{tikz}
\usepackage{subfigure}

\usepackage{diagbox}
\usepackage{booktabs}
\usepackage{multirow}
\usepackage{makecell}
\usepackage{amsfonts}
\usepackage{amssymb}
\usepackage{graphicx}
\usepackage{dcolumn}
\usepackage{bm}
\usepackage{amsmath}
\usepackage{graphicx}
\usepackage{subfigure}
\usepackage{float}
\usepackage{color}

\begin{document}

\preprint{APS/123-QED}
\preprint{This line only printed with preprint option}

\title{Isolated extended states and anomalous critical behavior in the generalized SSH model}

\author{Jia-Ming Zhang}

\affiliation {School of Physics, University of Science and Technology of China, Hefei 230026, China}

\date{\today}

\begin{abstract}
We investigate the localization properties of a generalized SSH model. Numerical and analytical results indicate the emergence of extended states protected by unbounded hopping in this model. Moreover, this protection effect is disrupted by the appearance of generalized incommensurate zeros, causing the extended phase in the system to transition into a multifractal phase. However, at the boundaries of the phase region, we still observe the existence of extended states. These extended states coincide with multifractality-enriched mobility edges, separating the multifractal phase from the localized phase. Further analysis reveals that this extended states originates from the band edge states of SSH model. In addition, these isolated extended states also influence eigenstates with nearby energies, giving rise to an anomalous extended-to-multifractal critical transition. These findings not only enrich the behavioral repertoire of eigenstates at critical points, but also offer new insights for further understanding Anderson localization and the induction of multifractal phases.
\end{abstract}

\maketitle

\section{Introduction}

In 1958, inspired by electron spin resonance experiments on silicon conducted by G. Feher and E.A. Gere, Anderson established his renowned theory of localization to characterize the metal-insulator phase transition induced by disorder~\cite{PWAnderson1958, GFeher1959a, GFeher1959b}. Subsequently established scaling theory~\cite{EAbrahams1979, PALee1985, BHetenyi2021} indicates that in low-dimensional (1D, 2D) systems, any amount of disorder leads to complete localization. Only in three-dimensional and higher-dimensional systems can extended states and localized states coexist, with an energy threshold—termed the mobility edges(MEs)—emerging to distinguish between them~\cite{FEvers2008, ALagendijk2009, NMott1987}. MEs possess significant application value, such as being utilized to realize thermoelectric devices with strong thermoelectric effects~\cite{RWhitney2014, YamamotoK2017, CChiaracane2020}.

Quasiperiodic systems, which lie between disordered systems and periodic systems, can not only induce mobility edges under low-dimensional conditions but also possess a concise mathematical form that is amenable to analytical treatment compared to three-dimensional disordered systems. They serve as an important platform for studying Anderson localization in low-dimensional systems~\cite{DJThouless1988, MKohmoto1983, MKohmoto2008, XCai2013, GRoati2008, YLahini2009, DTanese2014, HPLuschen2018, FAAn2018,FAAn2021, YWang2022a, HLi2023}. The AA model is a paradigmatic quasiperiodic system, renowned for its self-duality~\cite{SAubry1980, PGHarper1955, MGonalves2022}. Specifically, in the AA model, there exists a critical quasiperiodic disorder strength: below this critical value, only the extended phase exists in the system; above it, only the localized phase exists. Exactly at the critical point, the system exhibits critical features, which are referred to as the multifractal phase. Although the AA model does not feature mobility edges, by using it as a framework and modifying the structure of its quasiperiodic potential~\cite{SDSarma1988, SDSarma1990, ZLu2022, YWang2020a, SGaneshan2015, HYao2019, XLi2020, TLiu2022, XPLi2016, XLi2017, BFZhu2023, EWLiang2023, SLZhu2013} and hopping terms~\cite{JBiddle2010, JBiddle2011, XDeng2019, XXia2022, MGon2023, XCZhou2023a}, or extending it to the non-Hermitian regime~\cite{YJZhao2025, SZLi2024a, SZLi2024b, GJLiu2024, SZLi2024c, SLJiang2023, DWZhang2020a, DWZhang2020b, HJiang2019, JLDong2025}, a wealth of phenomena including MEs can be induced.

Considerable evidence suggests that quasiperiodic systems can host a third type of phase that is neither extended nor localized, namely the multifractal phase. This phase derives its name from the multifractal characteristics of its eigenstate wavefunctions and exhibits dynamical and spectral statistical properties that are distinctly different from those of the other two phases~\cite{MGon2023, HLi2023, YHatsugai1990, JHHan1994, YTakada2004, FLiu2015, JWang2016}. Recent studies suggest that this peculiar phase may be related to enhancing superconducting critical temperatures~\cite{JMayoh2015, MVFeigelman2007, MVFeigelman2010, ZFan2021, XZhang}, sparking significant interest in multifractal phases and multifractality-enriched mobility edges(MMEs)~\cite{SZLi2025}.

Within the framework of the AA model, there are primarily two mechanisms for inducing multifractal phase regions or MMEs. The first involves introducing an unbounded quasiperiodic potential, as indicated by the Simon-Spencer theorem~\cite{BSimon1989, SLonghi2023}. The second involves introducing incommensurately distributed zeros(IDZs), as indicated by Zhou's theorem~\cite{XCZhou2023a, XCZhou2025}. The common feature of these two mechanisms is the introduction of quasiperiodically distributed sites that impede wavefunction propagation, effectively cutting the chain into several subchains and precluding the existence of absolutely continuous(AC) spectra~\cite{Avila2017}. Additionally, a recent study indicates that due to the mutual mapping between potential terms and hopping terms in quasiperiodic systems, there actually exists a hidden structural connection between the two aforementioned induction mechanisms. Moreover, this structural connection can be utilized to achieve further manipulation of the induction mechanisms, such as by employing properly tailored hopping terms to protect extended states in systems with unbounded potential(UP)~\cite{JMZhang2025, HTHu2025}.

However, the present study breaks through the above paradigm, the main findings of this manusciprt are as follows: {\it We not only observe isolated extended states in a system that simultaneously possesses both UP and IDZs, but also find that these special states induce anomalous critical behavior.}

The rest of this manuscript is organized as follows:
In Section~\ref{Sec2}, we introduce the quasiperiodically modulated generalized SSH model and define the key physical quantities required for subsequent analysis. Section~\ref{Sec3} presents the main results obtained from numerical calculations and analytical derivations. The isolated extended states and anomalous critical behavior will be discussed in Section~\ref{Sec4}. Finally, we summarize our conclusions in Section~\ref{Sec5}.

\section{Model and key observables}\label{Sec2}

We obtain our model by introducing unbounded quasiperiodic disorder into both the potential terms and the intra-cell hopping terms of the standard SSH Hamiltonian~\cite{WPSu1979}. The resulting Hamiltonian is given by
\begin{equation}\label{E1}
H=\sum_{j=1}^{L/2}(t_ja_{j}b_{j}^\dag+Jb_{j}a_{j+1}^\dag+H.c.)+\sum_{j=1}^{L/2}V_j(b_jb_j^\dag+a_ja_j^\dag)
\end{equation}
where
\begin{equation}\label{E2}
V_j=\frac{\lambda}{\cos(2\pi\alpha j+\theta)}~~\&~~t_j=\frac{\lambda}{\cos(2\pi\alpha j+\theta)}+t
\end{equation}
with $a_{j}^{\dagger}$ and $b_{j}^{\dagger}$ ($a_{j}$ and $b_{j}$) denoting creation (annihilation) operators on $a$ and $b$ sublattices of the $j$-th site, respectively. $\lambda$ represents the strength of the quasiperiodic disorder introduced into the system, $t_j$ and $J$ correspond to the strength of intra- and extra-hopping of unit cells. $\alpha$ represents the quasiperiodic parameter, which is taken as the golden ratio $\frac{\sqrt{5}-1}{2}$. $\theta$ denotes the random phase, without loss of generality, is set to 0 in subsequent calculations. The system size is denoted as $L$.

Next, we introduce the key physical quantities used to characterize the localization properties of the system. The fractal dimension can be defined from the scaling of the moments of the wavefunction. Let the q weight of the wavefunction be defined by
\begin{equation}\label{E3}
\xi_{q}(\beta)=\sum^L_{j=1}\frac{\vert\psi_j(\beta)\vert^{2q}}{(\vert\psi_j(\beta)\vert^{2})^q}\varpropto L^{-\Gamma_q(q-1)},
\end{equation}
where $\vert\psi_j(\beta)\rangle=\sum_j\psi_j(\beta)\vert j\rangle$ for the $\beta$th eigenstate, and $\Gamma_q$ represents the fractal dimensions~\cite{HYao2019, XDeng2019, YWang2020a, SZLi2024c, AJ2021,YWang2022b}. This quantity can be used to numerically distinguish among three different types of eigenstates. If the wavefunction corresponds to an extended (localized) state, then $\Gamma_q=1(0)$ holds for all $q$ values. If the wavefunction is multifractal, $\Gamma_q$ lies between 0 and 1 and varies with $q$~\cite{AJ2021}. Here, we focus on the case $q=2$ to distinguish among the three different types of eigenstates, which is useful for further calculations. In this case, the corresponding quantity $\xi_{2}$ represents the inverse participation ratio (IPR). Then, subscripts omitted, the fractal dimension at finite size can be defined as
\begin{equation}\label{E4}
\Gamma(\beta)=-\lim_{L\rightarrow\infty}\frac{\ln\xi_{2}(\beta)}{\ln L}.
\end{equation}

By examining the scaling behavior of $\Gamma$ as it approaches the thermodynamic limit, specifically tending toward 1, 0, or a value between 0 and 1, one can distinguish between extended, localized, and multifractal states respectively. These results can be obtained through finite-size extrapolation~\cite{YWang2022b}. Furthermore, to compute the average fractal dimension within a mixed phase, we can define $N[\mathcal{R}]$ as the number of indices within a certain region $\mathcal{R}$ of the $\beta$ value range. Then, the average fractal dimension for a specific region within a mixed phase can be expressed as
\begin{equation}\label{E5}
\overline\Gamma=\frac{1}{N[\mathcal{R}]}\sum_{\beta\in\mathcal{R}}\Gamma(\beta).
\end{equation}
For cases specifically referring to a particular component within a mixed phase, we denote the regions occupied by extended, localized, and multifractal states as $\mathcal{R}=Ext.,~Loc.,~MF.$, respectively.

In addition to the fractal dimension, there is another numerical method to distinguish among extended, localized, and multifractal states, namely the even-odd(odd-even) spectral statistics. This numerical approach originates from the degeneracy properties of the energy spectrum in the extended phase under periodic boundary conditions. Since the wavefunctions of extended states exist in the form of plane waves, their real and imaginary parts each correspond to an eigenstate, resulting in doubly degeneracy of the energies in the spectrum~\cite{SAubry1980}.

We can extract the doubly degeneracy feature by performing statistical processing on the raw spectrum. For instance, computing the differences between energies with odd and even indices in the ascendingly ordered spectrum, namely the even-odd (odd-even) level spacings. The sets are defined as:
\begin{equation}\label{E6}
\delta_n^{e-o,o-e}=E_{2n}-E_{2n-1},E_{2n+1}-E_{2n}.
\end{equation}
Here, $n=1,2,3,\dots,n_{max}$~\cite{APadhan2022, YZhang2022, MSarkar2021, XDeng2019, RQi2023}, denotes the index of the even-odd (odd-even) energy level spacings arranged in ascending order. It is straightforward to see that if doubly degeneracy exists in the spectrum, a noticeable gap will appear between the sets $\delta_n^{e-o}$ and $\delta_n^{o-e}$. In other words, the emergence of a gap between the two sets signals the appearance of the extended phase.

Although this gap tends to close in the thermodynamic limit, the degree of closure differs for different types of phases. As a result, the logarithms of the two sets can still indicate the doubly degeneracy feature of the spectrum as the thermodynamic limit is approached. Specifically, as the system approaches the thermodynamic limit, for the extended phase, a noticeable gap emerges between sets $\ln\delta_n^{e-o}$ and $\ln\delta_n^{o-e}$; for the localized phase, the gap disappears; and for the multifractal phase, due to its critical behavior intermediate between the extended and localized phases, the two sets exhibit a scattered distribution distinctly different from the previous two characteristics~\cite{XDeng2019}.

\section{The numerical and analytical results}\label{Sec3}
\begin{figure}[htbp]
\centering
\includegraphics[width=8.5cm]{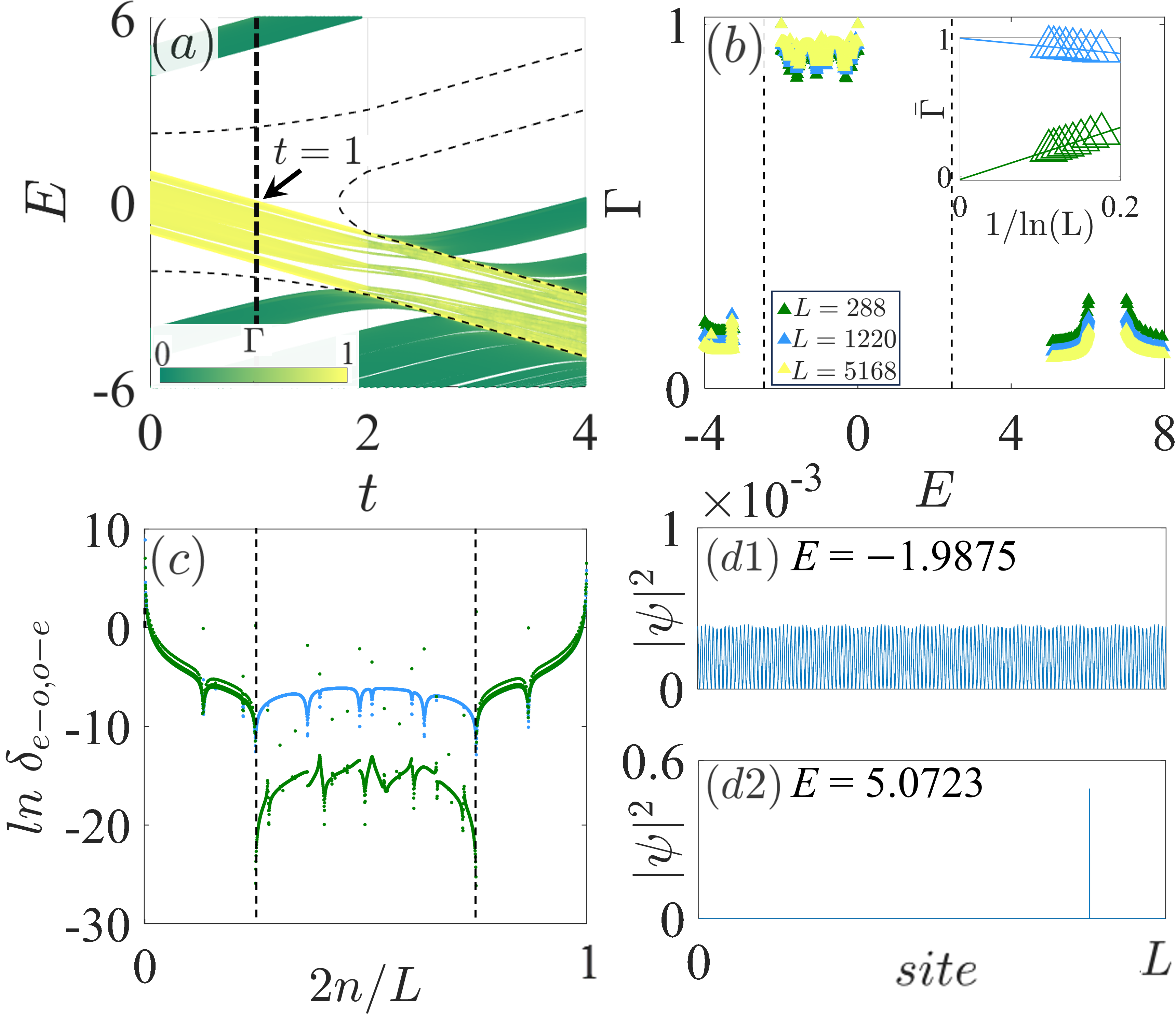}
\caption{(a) Phase diagram of the fractal dimension with $t$ and $E$ as parameters, where $\lambda=2$ and $L=1220$, the thick black dashed line marks $t=1$. (b) Cross sections of (a) at $t=1$ for different system sizes. The inset shows the variation trend of the fractal dimension as it approaches the thermodynamic limit for the extended(blue) and localized(green) regions separated by the MEs, obtained via finite-size extrapolation, with $L=$ 288, 466, 754, 1220, 1974, 3194, 5168 and 8362. (c) The level spacings $\ln\delta^{e-o}$ (blue) and $\ln\delta^{o-e}$ (green) at $t=1$, $\lambda=2$ for a system size $L=5168$. (d1) and (d2) display the wavefunction distributions in the extended region and the localized region for $t=1$ and $\lambda=2$, respectively. Throughout, we set $J=1$ as the energy unit, the MEs are marked with thin black dashed lines.}
\label{F1}
\end{figure}

Through numerical calculations, we first demonstrate the existence of the extended state protection effect in this model. The corresponding results are presented in Fig.~\ref{F1}.

Fig.~\ref{F1}(a) displays the phase diagram of the fractal dimension in the $t-E$ plane for $\lambda=2$. We first discuss the case where $t<\lambda$. When $t=1$, the results of the fractal dimension (see Fig.~\ref{F1}(b)) show that the spectrum of the system is divided into two regions. In these regions, the fractal dimension tends to 1 or 0 as the system size increases, indicating that only extended or localized components exist within the mixed phase. The inset presents a finite-size extrapolation for the two regions, showing that the average fractal dimension in the thermodynamic limit extrapolates to 1 or 0, which further confirms the conclusion that the MEs distinguishes only between extended and localized phases. Furthermore, the results of the even-odd (odd-even) spectral statistics also support this conclusion. From the results shown in  Fig.~\ref{F1}(c), it can be observed that either a noticeable gap appears between sets $\ln\delta^{e-o}$ and $\ln\delta^{o-e}$, or the gap closes. This indicates that only extended-localized mixed phases exist in the system, with no evidence of a multifractal phase. Thus, these findings provide a double verification consistent with the previous fractal dimension results. Finally, Fig.~\ref{F1}(d1) and (d2) display the spatial distributions of eigenstates extracted from the extended and localized regions, providing further strong evidence for the distinction between these two types of regions in the system.

\begin{figure}[htbp]
\centering
\includegraphics[width=8.5cm]{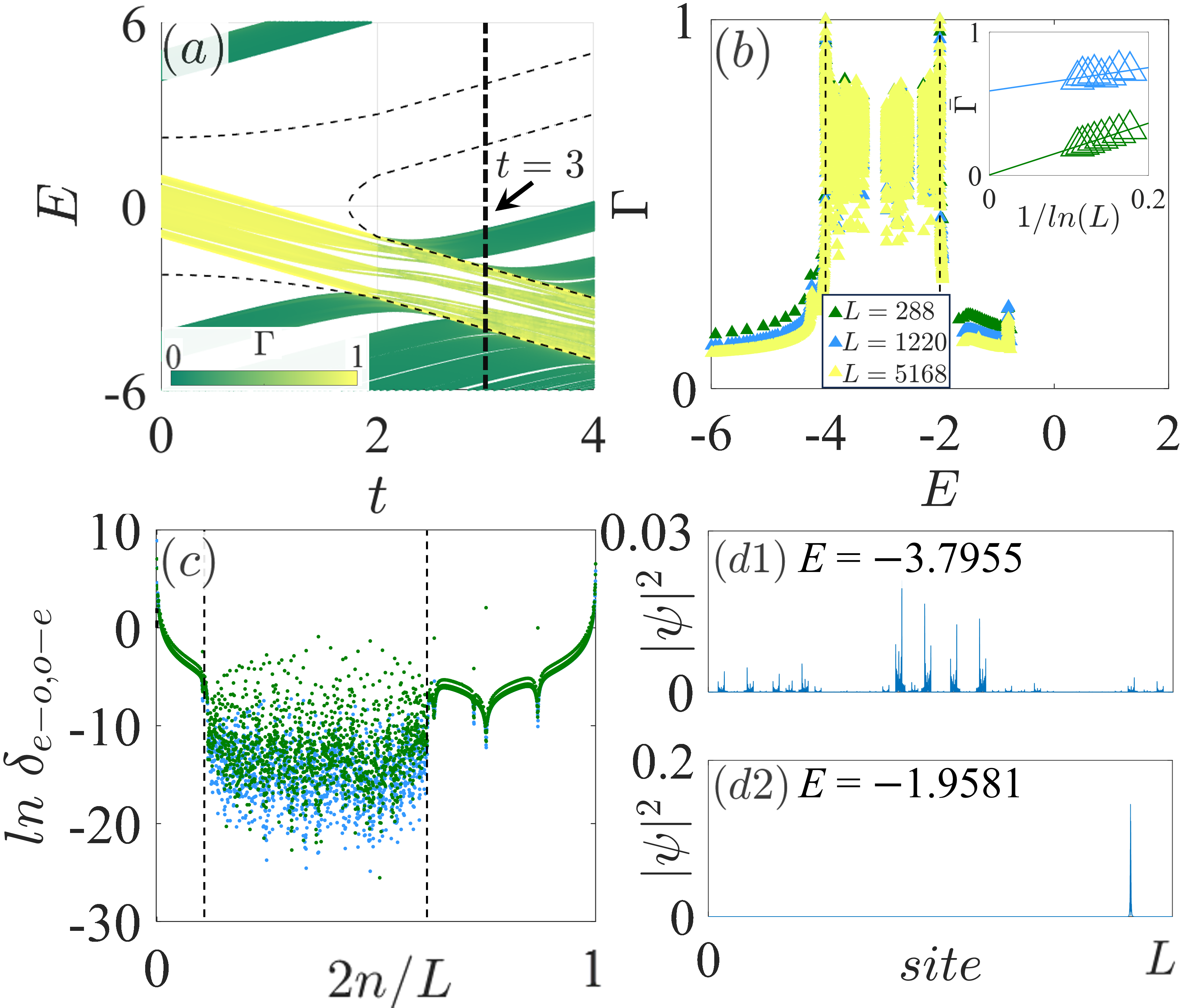}
\caption{(a) Phase diagram of the fractal dimension with $t$ and $E$ as parameters, where $\lambda=2$ and $L=1220$, the thick black dashed line marks $t=3$. (b) Cross sections of (a) at $t=3$ for different system sizes. The inset shows the variation trend of the fractal dimension as it approaches the thermodynamic limit for the multifractal(blue) and localized(green) regions separated by the MMEs, obtained via finite-size extrapolation, with $L=$ 288, 466, 754, 1220, 1974, 3194, 5168 and 8362. (c) The level spacings $\ln\delta^{e-o}$ (blue) and $\ln\delta^{o-e}$ (green) at $t=3$, $\lambda=2$ for a system size $L=5168$. (d1) and (d2) display the wavefunction distributions in the multifractal region and the localized region for $t=3$ and $\lambda=2$, respectively. Throughout, we set $J=1$ as the energy unit, the MEs are marked with thin black dashed lines.}
\label{F2}
\end{figure}

It is worth noting that the above results demonstrate the emergence of extended states, even though an unbounded potential exists in the system. Generally, the presence of an unbounded potential precludes the existence of AC spectra, leading to the absence of extended states and inducing the emergence of multifractal regions~\cite{BSimon1989, Avila2017}. The observed emergence of extended states indicates that an extended state protection effect occurs in the system, which results from the unbounded hopping overcoming the influence of the unbounded potential~\cite{JMZhang2025}. However, this effect is not always present. When $t\ge\lambda$, this protection effect disappears, and the MEs in the system transition into MMEs. Fig.~\ref{F2} presents the corresponding results for $t=3$.

Unlike the results in Fig.~\ref{F1}, although the fractal dimension results in Fig.~\ref{F2}(b) also show that the energy spectrum is divided into two regions, the average fractal dimensions of these two regions extrapolate to a value significantly between 0 and 1 and to 0, respectively, in the thermodynamic limit. This indicates that the two regions correspond to multifractal states and localized states, respectively, and are separated by MMEs. The even-odd(odd-even) spectral statistics yield consistent results. As shown in Fig.~\ref{F2}(c), either no gap appears between sets $\ln\delta^{e-o}$ and $\ln\delta^{o-e}$, or a scattering behavior emerges directly. This serves as another important piece of evidence for the existence of a multifractal-localized mixed phase in the system. Fig.~\ref{F2}(d1) and (d2) present the spatial distributions of wavefunctions in the two regions, providing further support for the above conclusions.

The mechanism underlying this phenomenon is quite clear. When $t\ge\lambda$, IDZs begin to appear in the hopping terms of the system. The presence of IDZs imposes additional disruption to the extended states, which cannot be overcome by the unbounded hopping. To support this conclusion, we provide the analytical solution of the model. From Eq.~\eqref{E1}, the transfer matrix expression within a unit cell can be obtained
\begin{equation}\label{E7}
\begin{split}
T_{j}(\theta)&=
\left(
\begin{array}{cc}
\frac{E-V_j}{J} & -\frac{t_j}{J} \\
1 & 0
\end{array}
\right)
\left(
\begin{array}{cc}
\frac{E-V_j}{t_j} & -\frac{J}{t_j} \\
1 & 0
\end{array}
\right)\\
&=
\left(
\begin{array}{cc}
\frac{(E-V_j)^2-t_j^2}{J}  & V_j-E \\
E-V_j  & -J
\end{array}
\right).
\end{split}
\end{equation}
By substituting the specific forms of $V_j$ and $t_j$, the above expression can be written as follows
\begin{equation}\label{E8}
T_j(\theta)=A_jB_j,
\end{equation}
where
\begin{equation}\label{E9}
\begin{split}
A_j&= \frac{1}{\lambda + tC_j}, \\
B_j&= 
\left(
\begin{array}{cc}
\frac{(E^{2}-t^2)C_j-2E\lambda-2t\lambda}{J} & -(EC_j-\lambda) \\
EC_j-\lambda  & -JC_j
\end{array}
\right),
\end{split}
\end{equation} 
with $C_j= \cos(2\pi\alpha j + \theta)$. The Lyapunov exponent is written as~\cite{SZLi2024a}
\begin{equation}\label{E10}
\gamma(E) = \lim_{L\to\infty} \frac{1}{L}\ln\left\|\prod_{j=1}^L T_j(\theta)\right\| = \gamma_A(E) + \gamma_B(E).
\end{equation} 
Using the ergodic theory~\cite{SLonghi2019}, we obtain
\begin{equation}\label{E11}
\begin{aligned}
\gamma_A(E) &= \lim_{L\to\infty} \frac{1}{L}\ln\left\|\prod_{j=1}^L \frac{1}{\lambda + tC_j}\right\| \\
&= \frac{1}{2\pi}\int_0^{2\pi} \ln\frac{1}{|\lambda + t\cos(\theta)|}d\theta \\
&= 
\begin{cases}
\ln\frac{2}{\lambda+\sqrt{\lambda^2-t^2}}, & t\le\lambda, \\
\ln\frac{2}{t}, & t>\lambda.
\end{cases}
\end{aligned}
\end{equation} 
For the $\gamma_B(E)$, we can apply the Avila's global theory of one-frequency analytical $ SL(2,\mathbb{C}) $ cocycle~\cite{{Avila2015}}. Under the case of large $\epsilon$ limit, by complexifying the phase as $\theta \to \theta + i\epsilon$, one can get
\begin{equation}\label{E11}
B_j(\epsilon) = \frac{e^{-i2\pi\alpha j }e^{|\epsilon|}}{2}
\begin{pmatrix}
 \frac{E^2-t^2}{J} & -E \\
E & -J
\end{pmatrix} + \mathcal{O}(1).
\end{equation} 
The corresponding Lyapunov exponent is given by
\begin{equation}\label{E13}
\begin{aligned}
\gamma_B(E) &= \lim_{L\to\infty} \frac{1}{L}\ln\left\|\prod_{j=1}^L B_j(\epsilon)\right\| \\
&= \ln\left|\frac{\kappa\pm\sqrt{-\kappa-4J^{2}t^2} }{4J}\right|+ \epsilon + \mathcal{O}(1),
\end{aligned}
\end{equation} 
where $\kappa=E^2-t^2-J^2$. According to Avila's global theory, $\gamma(E)$ is a convex piecewise linear function with respect to $\epsilon$. In large $\epsilon$ limit, the slope of $\gamma$ is always one, which means that $\mathcal{O}(1)$ term can be ignored. Since Lyapunov exponent's expression is an even function, the energy $E$ belongs to the spectrum of Hamiltonian~\eqref{E1}, we have
\begin{equation}\label{E14}
\gamma_B(E) = \max\left\{\ln\vert\frac{\kappa\pm\sqrt{-\kappa-4J^{2}t^2} }{4J}\vert+ \vert\epsilon\vert, 0\right\}.
\end{equation}
Since $\epsilon$ affects Lyapunov exponents is merely to increase its overall value without qualitatively changing, it is safe to drop the $\epsilon$ term from the final expression. Combined with $\gamma_A(E)$, and letting $\epsilon=0$, the Lyapunov exponents is
\begin{equation}\label{E15}
\gamma(E)=
\begin{cases}
\max\left\{\ln\vert\frac{\kappa\pm\sqrt{-\kappa-4J^{2}t^2} }{2J(\lambda+\sqrt{\lambda^2-t^2)}}\vert,0\right\}, & t\le\lambda \\
\max\left\{\ln\vert\frac{\kappa\pm\sqrt{-\kappa-4J^{2}t^2} }{2Jt}\vert,0\right\}, & t>\lambda
\end{cases}
\end{equation}
From the above expression, one can find that when the input value in natural logarithm function is greater than one, the Lyapunov exponent is larger than zero, which means $\ln|1|$ is the critical point. Based on this, one can obtain the analytical expression of the mobility edge, i.e.,
\begin{equation}\label{E16}
\begin{cases}
\vert\kappa\pm\sqrt{-\kappa-4J^{2}t^2}\pm2J(\lambda+\sqrt{\lambda^2-t^2})\vert=0, & \lambda \ge t,\\
\vert\kappa\pm\sqrt{-\kappa-4J^{2}t^2}\pm2Jt\vert=0, & \lambda < t.
\end{cases}
\end{equation}
The analytical results agree with numerical ones well (see black dashed line in Fig.~\ref{F1} and Fig.~\ref{F2}).

\section{Isolated extended states and anomalous critical behavior}\label{Sec4}

\begin{figure}[htbp]
\centering
\includegraphics[width=8.5cm]{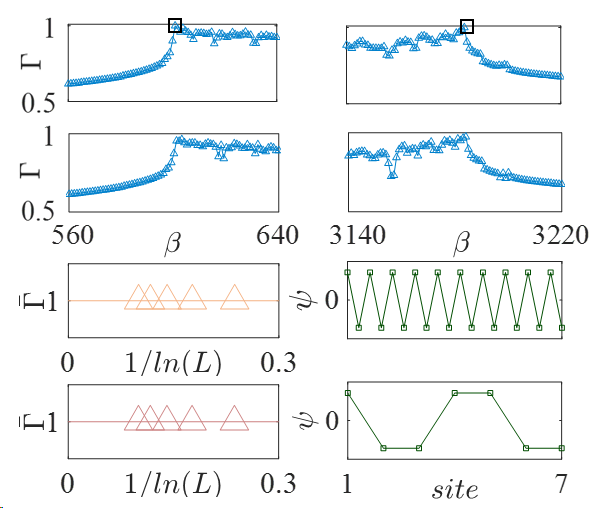}
\caption{(a) and (b): Detailed distribution of the fractal dimension under different boundary conditions. (c) Thermodynamic limit extrapolation of the fractal dimension for band edge states. (d) Wave functions of the band edge states.}
\label{F3}
\end{figure}

Now, let us turn our attention to the discussion of states at MMEs. Generally, the behavior of states at a critical point should lie between the two phases separated by that point. However, we find evidence of extended states at MMEs. The corresponding results are shown in Fig~\ref{F3}.


First, Fig~\ref{F3}(a) and (b) show the details of the fractal dimension near the MMEs. It can be observed that under periodic boundary conditions(PBC), a state with a fractal dimension of 1 appears at the MMEs, whereas this state disappears under open boundary conditions(OBC). The finite-size analysis in Fig~\ref{F3}(c) demonstrates that the existence of this isolated extended state under PBC is stable and unaffected by changes in system size, indicating that it is not a numerical artifact.

\begin{figure}[htbp]
\centering
\includegraphics[width=8.5cm]{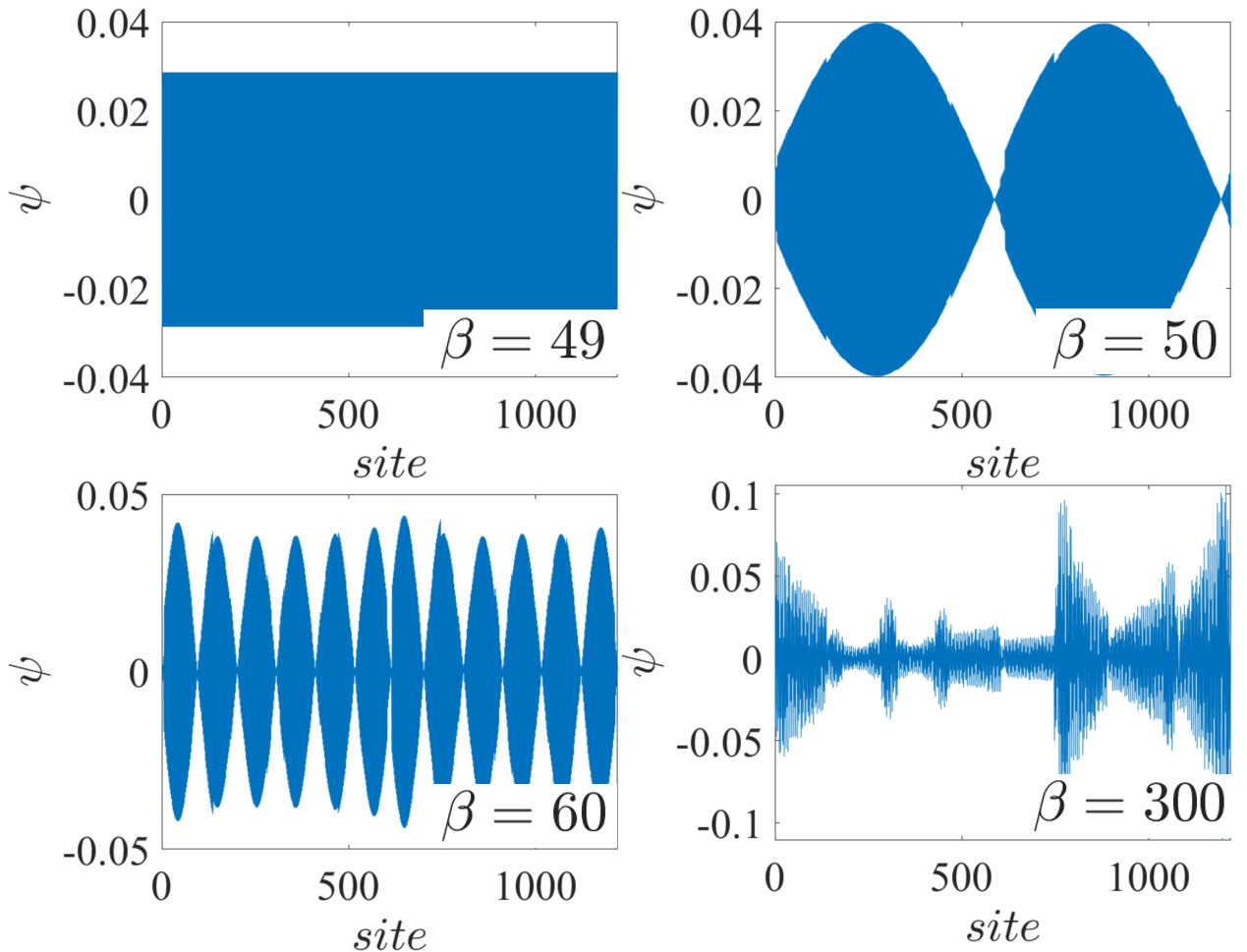}
\caption{Wave functions for different $\beta$ values under the same parameters.}
\label{F4}
\end{figure}

Furthermore, we uncover a critical behavior distinct from that of conventional MMEs. Specifically, conventional MMEs separate localized states from multifractal states, with states in their vicinity exhibiting a transition from localized to multifractal behavior. In contrast, we find that although the MMEs in our model separate localized states from multifractal states, their critical behavior corresponds to a transition from extended states to multifractal states, as shown in Fig~\ref{F4}.

We now provide an analytical explanation for the origin of these anomalous phenomena. Let us set $\lambda=0$, reducing Eq.~\eqref{E1} to the standard SSH model, and begin our derivation.
\begin{equation}
H_{SSH}=\sum_{j=1}^{L/2}[ta_{j}b_{j}^\dag+Jb_{j}a_{j+1}^\dag+H.c.]
\end{equation}
By applying a dual transformation, the Bloch Hamiltonian is obtained.
\begin{equation}
H_{SSH,k}=\sum_{k}[(t+Je^{ik})a_{k}b_{k}^\dag+H.c.]
\end{equation}
written in matrix form
\begin{equation}\label{Eq}
H_{SSH,k}(k)=
\left(
\begin{array}{cc}
0 & t+Je^{-ik} \\
t+Je^{ik} & 0
\end{array}
\right)
\end{equation}
the energy spectrum can be obtained as
\begin{equation}
E_{SSH}=\pm\sqrt{t^2+J^2+2Jt\cos{k}}
\end{equation}
and the corresponding eigenstates as
\begin{equation}
\vert\Psi_\pm\rangle=
\left(
\begin{array}{c}
1 \\
\pm e^{i\theta_k}
\end{array}
\right)
\end{equation}
where $\theta_k$ is the argument of $t+Je^{ik}$ on the complex plane. According to Bloch's theorem, the eigenwavefunction in real space takes the following form.
\begin{equation}
\vert\psi_\pm(k)\rangle=\frac{e^{ikj}}{\sqrt{L}}\sum_{j=1}^{L/2}[u_A(k)\vert j,A\rangle+u_B(k)\vert j,B\rangle]
\end{equation}
where $u_A(k)$ and $u_B(k)$ are given by the eigenvectors of Eq.\eqref{Eq}. We highlight four special states with $k=0$ or $k=\pi$, which will be beneficial for our further discussion(Tab.~\ref{Tab1}).
\begin{table}[tbhp]
\renewcommand{\arraystretch}{2}
	\centering
 	\caption{The four band edge states in the SSH model}
	\begin{tabular}{ c c c c}
		\hline\hline
		Energy    &             Intra-cell Phase      &             Inter-cell Phase       &    Mode
 \\  \hline
		$E=t+J$      & Opposite           & Same   & + + + + + +\dots   \\
		$E=t-J$      & Opposite          &       Opposite            &   + + - - + +\dots  \\
		$E=-(t-J)$      & Same          &       Opposite            &  + - - + + -\dots  \\
		$E=-(t+J)$      & Same           & Same   & + - + - + -\dots  \\
\hline\hline
	\end{tabular}
	\label{Tab1}
\end{table}

Now, we focus on the doped case.  Eq.\eqref{E1} can be expressed as
\begin{equation}\label{E1}
H=H_{SSH}+H_\Lambda
\end{equation}
where
\begin{equation}\label{E1}
H_\Lambda=\sum_{j=1}^{L/2}(V_ja_{j}b_{j}^\dag+H.c.)+\sum_{j=1}^{L/2}V_j(b_jb_j^\dag+a_ja_j^\dag)
\end{equation}
when $\theta_k=0$, $\vert\psi_-(k)\rangle$ is simultaneously an eigenstate of $H_{SSH}$ and $H_\Lambda$. Therefore, $\vert\psi_-(k)\rangle$ remains an eigenstate of $H$. For $\vert\psi_+(k)\rangle$, the above condition does not hold, and all eigenstates in this energy branch are completely disrupted by the impurities.

As $t$ increases, the range of $\theta_k$ rapidly converges to zero, and $\psi_-(k)$ can be approximately regarded as an eigenstate of $H_\Lambda$. The range of $k$ values, along with the effect of $\theta_k$ being equivalent to a small perturbation when $\theta_k=0$, implies that all in-gap states originate from extended band edge states subjected to weak impurity perturbations. This elucidates the origin of the critical behavior from extended to multifractal states.

\section{CONCLUSION}\label{Sec5}

In summary, we have investigated a quasiperiodically modulated SSH model. Both numerical and analytical results reveal the occurrence of the extended-state protection effect in the system, which is suppressed by the presence of  IDZs. However, even in the presence of both UP and IDZs, extended states are still observed, and these extended states coincide with the MMEs, giving rise to an anomalous extended-to-multifractal critical behavior at the MMEs. Further analysis indicates that this phenomenon originates from the band edge states inherited from the SSH model, with in-gap states effectively corresponding to perturbations of these band edge states. These findings not only deepen our understanding of eigenstate behavior near critical points but also offer new perspectives for further comprehending Anderson localization and multifractal induction.

\section{Acknowledgements.}
We thank Yun-Yan Chen, Shan-Zhong Li and Zhi Li for their insightful suggestions.

\appendix

\end{document}